\documentclass[prc,twocolumn,showpacs,preprintnumbers,superscriptaddress,floatfix,nofootinbib]{revtex4-1}
\usepackage{amsfonts}
\usepackage{graphicx,,booktabs}
\usepackage{amssymb}
\usepackage{amsmath,txfonts,ulem}
\usepackage{graphicx}
\usepackage{dcolumn}
\usepackage{color}
\usepackage{bm}
\usepackage{ulem}
\usepackage{multirow}
\usepackage[colorlinks,
            citecolor=blue,
            anchorcolor=green,
            menucolor=orange,
            linkcolor=red,
            filecolor=red,
            runcolor=pink,
            urlcolor=blue,
            frenchlinks=red]{hyperref}

\allowdisplaybreaks[4]

\begin{document}
\title{{{ The $\psi ^{(\ast)}p$ scattering length based on near-threshold  charmoniums photoproduction   }}}

\author{Xiao-Yun Wang}
\email{xywang@lut.edu.cn}
\affiliation{Department of physics, Lanzhou University of Technology,
Lanzhou 730050, China}
\affiliation{Lanzhou Center for Theoretical Physics, Key Laboratory of Theoretical Physics of Gansu Province, Lanzhou University, Lanzhou, Gansu 730000, China}

\author{Fancong Zeng}
\email{fczeng@yeah.net}
\affiliation{Department of physics, Lanzhou University of Technology,
Lanzhou 730050, China}

\author{Igor I. Strakovsky}
\email{igor@gwu.edu}
\affiliation{Institute for Nuclear Studies, Department of Physics,
The George Washington University, Washington, DC 20052, USA}

\date{\today}
\begin{abstract}
Under the framework of Vector Meson Dominance model,    the  value of scattering length can be expressed as a function of the ratio between total cross section $\sigma (W)$ and $R (W)$, where $R(W)$ is the ratio between final   momentum $|{\bf p}_3 |$ and initial momentum $|{\bf p}_1|$ and positively correlated with the center-of-mass energy.
 Based on the theoretical study of charmoniums  photoproduction within  two gluon exchange model and effective pomeron model,  we research  the  scattering lengths  of   vector  mesons and proton interaction in this work.
 Results show that the  scattering length $\left|\alpha_{J/\psi p}\right|$ obtained from the two models are close and basically in agreement with the theoretical prediction of  Strakovsky and co-workers.
 Additionally, we first calculate  the scattering length  of  $\psi(2S) $-proton interaction  in two gluon exchange model and effective pomeron model as  $ 1.31\pm 0.92$ am (1 am = $10^{-3}$ fm) and $ 3.24 \pm 0.63$ am, respectively. This is a little bit different from the   two   models    and requires precise measurements from subsequent experiments.
In short,  our results will provide a theoretical reference for future studies on characterizing the vector meson-proton scattering length.
\end{abstract}


\maketitle
\section{INTRODUCTION}

 The evaluation of the scattering lengths    may serve as a unique input for QCD-motivated models of vector meson-nucleon interactions \cite{Strakovsky:2014wja,Strakovsky:2020uqs,Strakovsky:2019bev,Strakovsky:2021vyk}.
The behavior of the near-threshold  cross section   is related to vector meson-proton ($V p$)  scattering length
\cite{Gell-Mann:1961jim}.
 Recently, scattering lengths  for $\omega p, \phi p, J/\psi p $ and  $ \Upsilon p$ reactions have been reported using the recent  photoproduction experiment data or quasi data \cite{Strakovsky:2014wja,Strakovsky:2020uqs,Strakovsky:2019bev,Strakovsky:2021vyk}. However, there are also several $Vp$ scattering lengths that have not been studied for various reasons, such as  one narrow vector meson: $D^*_0(2007)$, since  no charm conservation, the reaction $\gamma p \rightarrow     D^*_0(2007)p$ is impossible.   Moreover, we find that there are no good threshold measurements for  $\rho, K^*$  and $ D^*$ meson photoproduction.
As the excited  states of  $J/\psi$,
  it is of great interest to  estimate the scattering length of  vector  meson $\psi(2S) $-proton interaction.

On the  experimental side, the $J/\psi $  photoproduction off the proton was conducted  with   increasing precision over a large energy range \cite{GlueX:2019mkq,ZEUS:2002wfj,Binkley:1981kv,E687:1993hlm,H1:2013okq,ALICE:2014eof,LHCb:2013nqs,ALICE:2018oyo},
while the measurement of $\psi(2S)$ photoproduction data is very meager \cite{H1:2002yab,LHCb:2018rcm,Hentschinski:2020yfm}.
Considering  $\psi(2S)$ and $J/\psi$ have close mass, both have $c \overline{c}$ structure,  the same quantum numbers, spin and parity as photon which is $I^G(J^{PC})=0^-(1^{--})$,   it is reasonable to study them with the same physical models and parameters.
 In our previous works \cite{Wang:2022vhr},  two gluon exchange model is applied to systematically  analyze the  $J/\psi$ and  $\psi(2S)$  photoproduction data from   threshold to medium energy (near  $ 400$ GeV) \cite{GlueX:2019mkq,ZEUS:2002wfj,Binkley:1981kv,E687:1993hlm,H1:2013okq,ALICE:2014eof,LHCb:2013nqs,ALICE:2018oyo}.
In the literatures  \cite{Winney:2019edt,Albaladejo:2020tzt}, the pomeron exchange process is  considered to explain the photoproduction of   charmoniums by JPAC collaboration.
The numerical results from two gluon exchange model and effective pomeron model  are both  in agreement with $J/\psi$ experimental data, while the predicted photoproduction of  $\psi(2S)$ in the two models   at the threshold is basically consistent \cite{Wang:2022vhr}.
In this paper,
we will use the  predicted  $\psi(2S)$ cross    section data under the framework of  the two models to extract the  scattering lengths  $\left|\alpha_{\psi(2S) p}\right|$.

Many literatures  give their  scattering length results for proton interacts with vector meson   by different methods.    From a global fit to both differential and total cross section data,  one work \cite{Gryniuk:2016mpk} extracted the scattering length $\left|\alpha_{J/\psi  p}\right|$  is $0.046 \text{ fm}$  from $d\sigma/dt(s_{thr},t$=$0)$.
 Ref. \cite{Kaidalov:1992hd}  discussed the  charmoniums
bound states in nuclei  from  the method of multipole expansion and low-energy QCD theorems.
 Provided in the Vector Meson Dominance (VMD) model \cite{ Titov:2007xb,Kroll:1967it},
Strakovsky and co-workers  estimated the $J/\psi$-nucleon scattering length  $\left|\alpha_{J/\psi p}\right|=3.08$ am
(1 am = $10^{-3}$ fm)  by fitting the recent GlueX total cross section data \cite{Strakovsky:2019bev}.
To avoid additional uncertainty when  extrapolating the differential cross section to the non-physical point $t$=0, the approach of ref. \cite{Gryniuk:2016mpk} is not adopted.
Compared to other methods, the VMD model  does not
 contain free parameters   in the process from  $\gamma p$ to  $V p$ reaction. It is superior for us to obtain  qualitative estimates when extracting scattering lengths  $\left|\alpha_{V p}\right|$.
The scattering length $\left|\alpha_{J/\psi p}\right| $  obtained by VMD model is smaller than other references.
And this small value can be attributed to   the size of   the ``Young"   vector meson   is smaller than that of the “old" one participating in the elastic $Vp \rightarrow  Vp$ scattering, because   $c \bar{c}$ pair  lacks sufficient time to form the complete wave function of the vector meson \cite{Strakovsky:2021vyk}.

 The measurement  of  near-threshold $ \psi(2S) $ photoproduction   will give access to a variety of interesting physics aspects, e.g., trace anomaly, pentaquarks, cusp effects, vector-meson-nucleon scattering length  and so  on \cite{Lutz:2001mi,Titov:2007xb}.
Nowadays, the   Electron Ion Colliders (EIC)
are   proposed to be built for probing the deepest structure inside the hadron  \cite{Accardi:2012qut}.
Meanwhile, the opportunities for Chinese EIC are now under discussion   and will be an important and interesting future machine to
collect $\psi^{(*)}$ data \cite{Anderle:2021wcy}.
Relevant measurements will advance our understanding of QCD which governs the properties of hadrons and the interactions involving hadrons.

The paper is organized as follows.
The formulas of effective pomeron model, two gluon exchange model   and an  expression for scattering length $\left|\alpha_{V p}\right|$
are provided in Sec.\ref{sec:formalism}.
Next section, we show the numerical result on the explanations of the current experimental data of $J/\psi$. Through the study of scattering length $\left|\alpha_{J/\psi p}\right|$, the reliability of the two models and VMD method are determined. Then the  results  and discussion   of $\left|\alpha_{\psi(2S) p}\right|$ are obtained in Sec. \ref{sec:results}. A  summary is given in  Sec. \ref{sec:summary}.

\section{  FORMALISM}\label{sec:formalism}
In quark-interchange mechanisms,
some  light quarks (such as $\pi$, $\rho$, $\Delta$, etc.) are strongly suppressed   in the heavy quarkonium photoproduction. So the channels are  always dominated by two gluon exchange mechanisms or pomeron exchange contributions.
Actually,  the two gluon exchange and   pomeron contributions both reproduce   well the existing data \cite{Laget:1994ba}.

 \subsection{ effective pomeron model }
 \begin{figure}[h!]
 \center
 \includegraphics[scale=0.43]{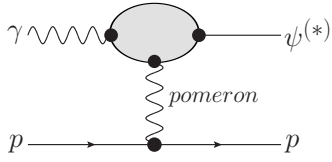}
 \caption{The schematic Feynman diagram of the  effective pomeron model  for  $ J/\psi $ or $ \psi(2S) $ production.}
 \label{fig:JPAC-model}
 \end{figure}
 The effective  $t-$channel pomeron exchange contribution is shown in Fig.\ref{fig:JPAC-model}.
Proposed by JPAC collaboration, an effective  pomeron exchange model is used to have a systematical analysis   for $J/\psi$ and  $\psi(2S)$ photoproduction   \cite{Albaladejo:2020tzt}.
The differential cross section for  $\gamma p\rightarrow  \psi ^{(\ast)}p$ reaction takes the form  \cite{Winney:2019edt}
\begin{align}
\label{eq:jpac}
\frac{d \sigma}{d t}=\frac{4 \pi \alpha}{64 \pi W^2 p_i^2}  \sum_{\lambda_{\gamma}, \lambda_p, \lambda_{\psi ^{(\ast)} }, \lambda_{p'}    }  \frac{1}{4} \left|\left<\lambda_{\psi ^{(\ast)}  } \lambda_{p'}|T| \lambda_{\gamma}  \lambda_p \right>\right|^2.
\end{align}
The amplitude  in low energy regions is given by \cite{Close:1999bi,Winney:2019edt}
\begin{equation}
\begin{aligned}
\left\langle\lambda_{\psi ^{(\ast)}} \lambda_{p^{\prime}}\left|T_{P}\right| \lambda_{\gamma} \lambda_{p}\right\rangle=& F(W^2, t) \bar{u}\left(p_{f}, \lambda_{p^{\prime}}\right) \gamma_{\mu} u\left(p_{i}, \lambda_{p}\right) \\
& \times\left[\varepsilon^{\mu}\left(p_{\gamma}, \lambda_{\gamma}\right) q^{\nu}-\varepsilon^{\nu}\left(p_{\gamma}, \lambda_{\gamma}\right) q^{\mu}\right] \\
& \times \varepsilon_{\nu}^{*}\left(p_{\psi ^{(\ast)}}, \lambda_{\psi ^{(\ast)}}\right) .
\end{aligned}
\end{equation}
Here,
\begin{equation}\label{eq:F}
 F(W^2,t)=i A_{\psi ^{(\ast)}} \left(\frac{W^2-W^2_{thr}}{W^2_0}\right)^{\alpha(t)} \frac{e^{b_0(t-t_{min})}}{W^2},
\end{equation}
 $\alpha(t)=\alpha_0+\alpha't$ is the pomeron trajectory \cite{V. N. Gribov}. $W_0=1 $ GeV  is the energy scale parameter. $W_{thr}= M_p+M_{\psi ^{(\ast)}} $ is the energy threshold.

 \subsection{two gluon exchange model}

 \begin{figure}[h!]
\center
\includegraphics[scale=0.38]{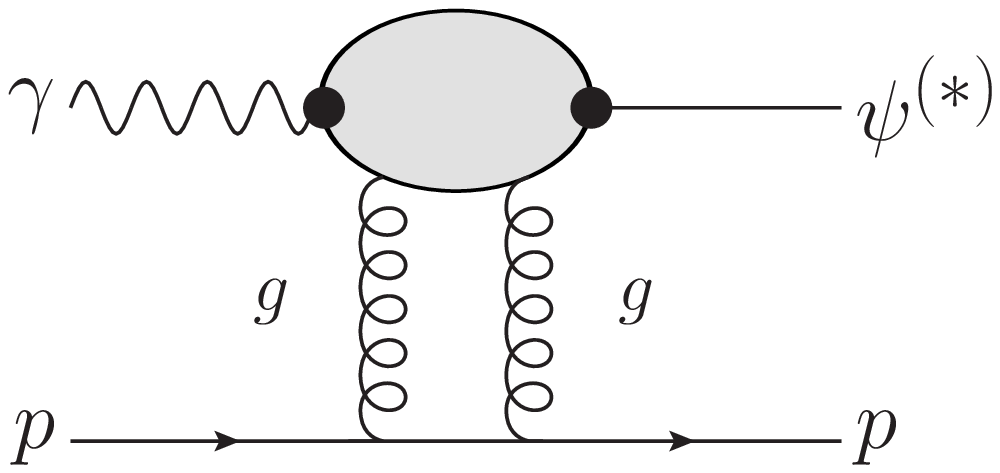}
\includegraphics[scale=0.38]{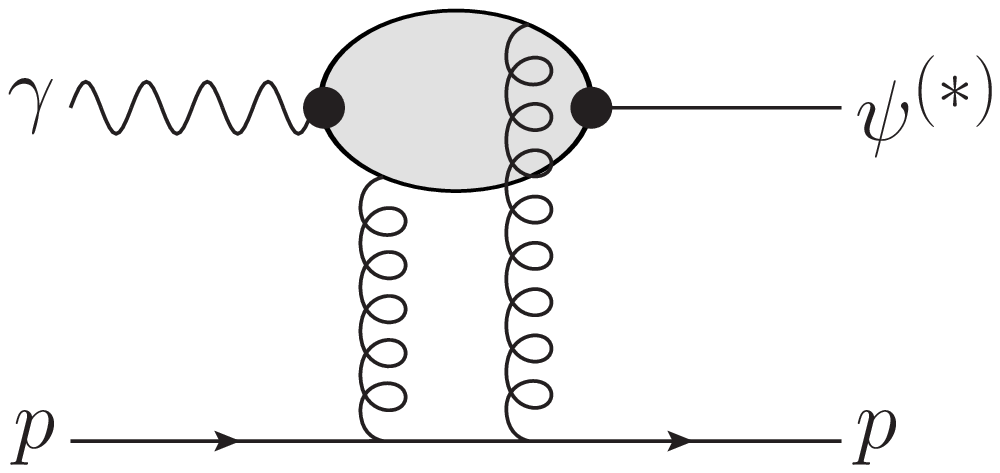}
\caption{The schematic Feynman diagrams of the two gluon exchange  model for  $ J/\psi $ or $ \psi(2S) $  production.}
\label{fig:TGE-model}
\end{figure}
The picture of the double gluon exchange between the nucleon state and the quark-antiquark pair  is illustrated in Fig.\ref{fig:TGE-model}.
 The   differential cross section of $ \psi ^{(\ast)}$  photoproduction is  given as \cite{Sibirtsev:2004ca,Zeng:2020coc}

\begin{align}
\label{eq:diff-crosssection}
\frac{d \sigma}{d t}=\frac{\pi^{3} \Gamma^{\psi ^{(\ast)}}_{e^{+}e^{-}} \alpha_{s}}{6 \alpha m_{q}^{5}}\left[x g\left(x, m_{\psi ^{(\ast)}}^{2}\right)\right]^{2} \exp (b^{\psi ^{(\ast)}} \cdot t),
\end{align}
where $x=m_{\psi ^{(\ast)} }^2/W^2$;
    $\alpha_{s}=0.5 $ is the strong  coupling constant \cite{Xu:2020uaa}; $\alpha$ is the fine-structure constant; $m_q$   is  the mass   of charm quark;
$\Gamma^{\psi ^{(\ast)}}_{e^+e^-} $ is the radiative decay \cite{Tanabashi:2018oca}.
$xg\left(x, m_{\psi ^{(\ast)}}^{2}\right)$ defines the gluon distribution function at $Q^2=m_{\psi ^{(\ast)} }^{2}$, which is parameterized using a simple function form $xg\left(x, m_{\psi ^{(\ast)}}^{2}\right)=A_0 x^{A_1}(1-x)^{A_2}$ \cite{Pumplin:2002vw}.
    The exponential slope parameter  $b$  can
use the standard form based on the Regge phenomenology \cite{Hentschinski:2020yfm,Cepila:2019skb} for $\psi(2S)$ and similar form \cite{Zeng:2020coc} for $J/\psi$.

The total cross section is obtained by integrating the differential cross section (Eq. (\ref{eq:jpac}) or (\ref{eq:diff-crosssection}))
over the allowed kinematical range from $t_{min}$ to $ t_{max}$,
  the total cross section  can be written as,
\begin{align}\label{eq:total}
\sigma=\int_{t_{min}}^{t_{max}}  dt  ~\frac{d \sigma}{d t}  ,
\end{align}
 here, the limiting values $t_{min}$ and  $t_{max}$ are
\begin{align}
t_{max}( t_{min})=m_1^2+m_3^2-2E_1 E_3  \pm   2 |{\bf p}_{1 }||{\bf p}_{3}|  .
\end{align}
The energies and momenta of the  photon and meson in the center-of-mass (c.m.) frame are

\begin{align}
|{\bf p}_{1 }|= \frac{1}{2W} \sqrt{W^4-2 (m_1^2+m_2^2) W^2 +(m_1^2-m_2^2)^2 }  ;
\end{align}

\begin{align}
|{\bf p}_{3}|= \frac{1}{2W} \sqrt{W^4-2 (m_3^2+m_4^2)  W^2 +(m_3^2-m_4^2)^2}  ;
\end{align}

\begin{align}
 E_1=\sqrt{{\bf |p_1|}^2+m_1^2},~~~ E_3=\sqrt{{\bf| p_3|}^2+m_3^2}.
\end{align}

\subsection{VMD model  and scattering length $\left|\alpha_{V p}\right|$}

The ratio between the initial c.m. momentum  and the final  momentum  $R(W)$ is used as
 \begin{align}
R(W)=\frac{|{\bf p}_3 |}{|{\bf p}_1| }.
\end{align}
Note that, $ R(W)$ has a range of $R(W)\in[0,1)$ and is positively correlated with the c.m. energy.    $R(W)\rightarrow 0$ respects  $W\rightarrow W_{thr}$ while $R(W)\rightarrow 1$ respects  $W\rightarrow \infty $.
The ratio of    total cross section $\sigma(R)$  and $R$ is given by
 \begin{align}\label{eq:a1}
  \frac{ \sigma(R)}{ R} =a_1(R),
\end{align}
where the function $a_1(R)$ can be described by  Eq. (\ref{eq:jpac}), (\ref{eq:diff-crosssection})  and (\ref{eq:total}).

The $V p$ scattering length  is  related to  the near-threshold  photoproduction of  vector mesons. In this paper,
the    VMD  model is used to connect the reaction
$\gamma p \rightarrow Vp$ and $ Vp \rightarrow Vp$.
 Applying the   effective VMD approach,
the near-threshold cross section during the elastic scattering  processes  becomes \cite{Titov:2007xb}
\begin{equation}\label{eq:sigma}
\begin{aligned}
\left. \sigma \right|_{thr}\left(R\right)  &= \left.\frac{|{\bf p}_3 |}{|{\bf p}_1| }   \cdot \frac{4  \alpha\pi^2}{g_{V}^{2}} \cdot \frac{d \sigma^{V p \rightarrow V p}}{d \Omega}\right|_{thr} \\
&= R \cdot \frac{4  \alpha\pi^2}{g_{V}^{2}}   \cdot\left|\alpha_{V p}\right|^{2},
\end{aligned}
\end{equation}
here
the VMD coupling constant  $g_{V}$   is  deduced from the  leptonic  decay width $\Gamma^{V}_{e^+e^-}$ as \cite{Strakovsky:2019bev}
\begin{align}\label{eq:gV}
g_{V}=\sqrt{\frac{\pi \alpha^2 	M_V}{3   \Gamma^{V}_{e^+e^-} }}.
\end{align}
Combining    Eq.(\ref{eq:a1}), (\ref{eq:sigma}) and  (\ref{eq:gV}), scattering length  $\left|\alpha_{V p}\right|$ is given as
\begin{align}\label{eq:alpha}
\left|\alpha_{Vp}\right| = \frac{g_{V}}{2\pi} \sqrt{\frac{ a_1(R)}{\alpha}} .
\end{align}

The scattering length  $\left|\alpha_{V p}\right|$ can also be expressed by   differential photoproduction cross section.
When the c.m. energy $W$ approaches the threshold,   the total cross section is related to  the differential cross section as \cite{Pentchev:2020kao}
\begin{align}\label{eq:2t}
\left. \sigma \right|_{thr}= |t_{max}- t_{min}|\left.   \frac{d  \sigma}{dt} \right|_{thr}=4  |{\bf p}_1 | \cdot |{\bf p}_3|     \left.  \frac{d  \sigma}{dt} \right|_{thr}.
\end{align}
 Combining Eq.   (\ref{eq:sigma}) and   (\ref{eq:2t}), we can obtain the relation between the  differential cross section of $\gamma p \rightarrow V p $ reaction   and  scattering length as
  \begin{align}
  \left.  \frac{d  \sigma}{dt} \right|_{thr}=\frac{\alpha \pi^2}{g_{V }^2 |{\bf p}_1|^2} \cdot \left|\alpha_{Vp}\right|^2.
\end{align}
If setting  $ \left.  \frac{d  \sigma}{dt} \right|_{thr}=b_1$,
the scattering length  $\left|\alpha_{V p}\right|$ is given as
   \begin{align}
\left|\alpha_{V p}\right|= \frac{g_{V}   |{\bf p}_1|  }{\pi} \sqrt{\frac{b_1}{\alpha}}.
\end{align}
Note that, $b_1$ and $|{\bf p}_1| $ must satisfy the  conditions that obtained from  the threshold.

\section{Results and discussions}
\label{sec:results}

\begin{figure}[b]
\begin{center}
\includegraphics[scale=0.4]{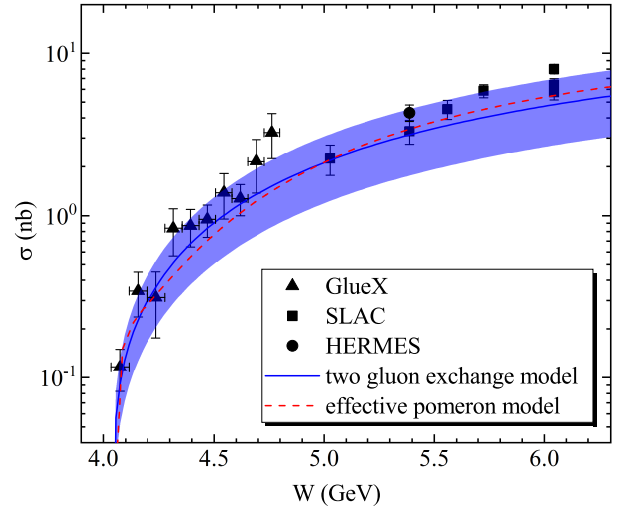}
\caption{The total  cross section of $\gamma p \rightarrow  J/\psi  p$   as a function of $W$. 	Blue-solid curve is obtained from two gluon exchange model. Red-dashed curve is obtained from effective pomeron model. Data are from Ref. \cite{GlueX:2019mkq,E687:1993hlm,Amarian:1999pi}.
}\label{fig:psi}
\end{center}
\end{figure}

\begin{figure}[htbp]
\begin{center}
\includegraphics[scale=0.4]{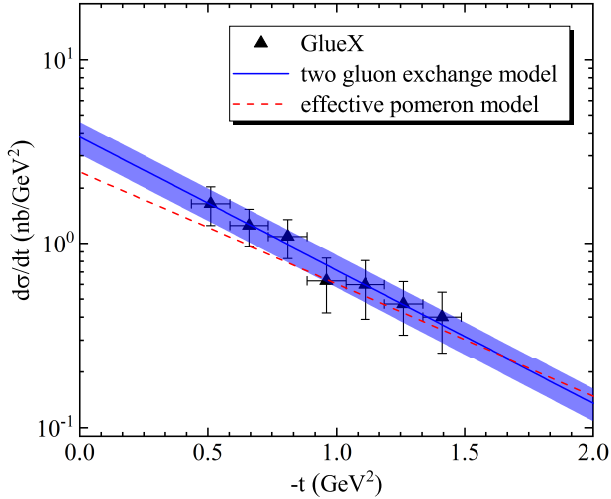}
\caption{The differential   cross section of $\gamma p \rightarrow  J/\psi  p$   as a function of $-t$. The curves have the same meaning as in Fig.\ref{fig:psi}.  Data are from Ref. \cite{GlueX:2019mkq}.
}\label{fig:psi-diff}
\end{center}
\end{figure}

In our previous work,
 the free parameters  $A_0, A_1$ and $A_2$ in two gluon exchange model were obtained by a global analysis of both the total cross section data below medium energy   \cite{GlueX:2019mkq,ZEUS:2002wfj,Binkley:1981kv,E687:1993hlm,H1:2013okq,ALICE:2014eof,LHCb:2013nqs,ALICE:2018oyo}  and the near-threshold (W = 4.58 GeV) differential  cross section data  of $J/\psi$ \cite{GlueX:2019mkq}.
JPAC collaboration determined the free parameters $A_{J/\psi}$, $\alpha_0$, $\alpha'$ and  $b_0 $   by fitting the  total cross section of $J/\psi$  experimental data \cite{GlueX:2019mkq,E687:1993hlm}.      The fitted  parameters from the two models  are listed in Tab. \ref{tab:GluonParameters} and  \ref{tab:JPAC}.
 The numerical results of $\gamma p \rightarrow J/ \psi  p$  from two gluon exchange model and      effective pomeron model     are shown in Fig.\ref{fig:psi} and  \ref{fig:psi-diff}, compared with GlueX ,  SLAC and HERMES experiments \cite{GlueX:2019mkq,E687:1993hlm,Amarian:1999pi}.
We perceive that the  two models are reliable to explain $J/ \psi  p$ photoproduction.

  \begin{table}
\centering
\caption{The fitted values of the parameters $A_0, A_1$ and $ A_2$  in two gluon exchange model \cite{Wang:2022vhr}.}
\begin{tabular}{ccccc}
\hline\hline\noalign{\smallskip}
  $A_0$ & $A_1$ & $ A_2$    \\
\noalign{\smallskip}\hline\noalign{\smallskip}
~~~~~ $0.228\pm 0.045$~~~~~ & ~~~~~$-0.218\pm 0.006$~~~~~ & ~~~~~$1.221\pm0.055$~~~~~   \\
\noalign{\smallskip}
\hline
\end{tabular}
\label{tab:GluonParameters}
\end{table}

\begin{table}
\centering
\caption{The fitted values of the parameters $A_{J/\psi}$, $\alpha_0$, $\alpha'  $ and  $b_0$ in   effective pomeron model \cite{Albaladejo:2020tzt}. }
\begin{tabular}{ccccc}
\hline\hline\noalign{\smallskip}
  $A_{J/\psi}$ & $\alpha_0$ & ~~~~~~$\alpha'$(GeV$^{-2})  $ ~~~~~~& ~~~~~~  $b_0 $(GeV$^{-2}) $ ~~~~~~ \\
\noalign{\smallskip}\hline\noalign{\smallskip}
 ~~~~~~ $0.38$~~~~~~ & ~~~~~~$0.94$~~~~~~ & ~~$0.36$~~ & ~~ 0.12~~ \\
\noalign{\smallskip}
\hline
\end{tabular}
\label{tab:JPAC}
\end{table}

\begin{figure}[htbp]
\begin{center}
\includegraphics[scale=0.4]{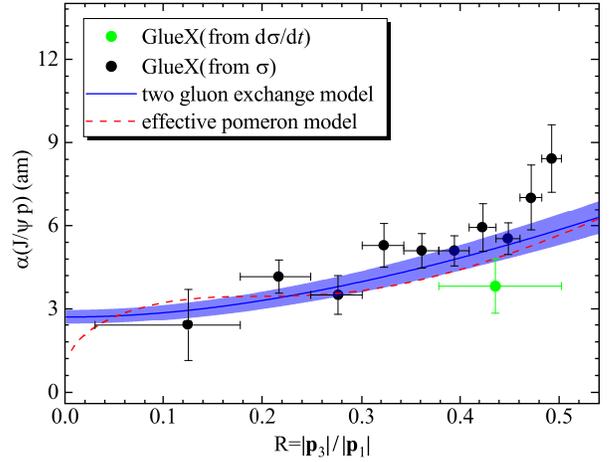}
\caption{  The obtained  scattering length  $\left|\alpha_{J/\psi p}\right|$ as a function of $R$. The black  circles show the results derived from the GlueX total cross section   data.  The green circle shows the results derived from the GlueX   differential cross  section data  \cite{Pentchev:2020kao}.    The curves have the same meaning as in Fig.\ref{fig:psi}.}
\label{fig:jpsi-R-total}
\end{center}
\end{figure}

According to the obtained total  cross section  from   models,  Eq. (\ref{eq:a1}) and  (\ref{eq:alpha}),   the  $J/\psi   p$ scattering length as a function of $R$ is  shown in Fig.\ref{fig:jpsi-R-total} (blue-solid curve).
Listed in Table  \ref{tab:jpsi-comparson},   the average value   of  scattering length $\left|\alpha_{ J/\psi  p}\right|$ is $ 3.85\pm0.96    \text{ am}$ from two gluon exchange model and  $ 3.75\pm0.84$ am from effective pomeron   model.  Here the range of $R$ is selected as $[0,0.5]$. Note that $R=0.5$ corresponds to  c.m. energy  $W= 4.79$ GeV, which represents a near-threshold energy.
We also compared our results with the phenomenological result    \cite{Pentchev:2020kao} and  odd-polynomial fitted result \cite{Strakovsky:2019bev}. Our results are in agreement with the above phenomenological results.

We also  extract  the scattering lengths from the  total cross section GlueX  data,
  as shown in the black  circles  in  Fig.\ref{fig:jpsi-R-total}.  Ref. \cite{Pentchev:2020kao}  obtained $\left|\alpha_{ J/\psi  p}\right|= 3.83\pm 0.98    \text{ am}$ (the green circle in  Fig.\ref{fig:jpsi-R-total})  derived from   the   differential cross  section GlueX data.
Although this   energy ($W=4.59$ GeV)  is a little far from  the threshold,  the value of  scattering length is close to  our  estimation  from two models and other works \cite{Pentchev:2020kao,Strakovsky:2019bev}.
Therefore,  the scattering lengths extracted from the differential cross section experimental data  may be more  advantageous, compared with the instability extracted from  total cross  section data.

\begin{table}
\centering
\caption{ Comparison of different determinations of  $\left|\alpha_{ J/\psi  p}\right|$ from different method.     }
\begin{tabular}{ccccc}
\hline\hline\noalign{\smallskip}
method  &      ~~   $\left|\alpha_{ J/\psi  p}\right|$ (am) ~~ \\
\noalign{\smallskip} \hline\noalign{\smallskip}
 odd-polynomial fit \cite{Strakovsky:2019bev}  &  $3.08 \pm 0.55$  \\
\noalign{\smallskip} \hline\noalign{\smallskip}
phenomenological result  \cite{Pentchev:2020kao} &    $3.64  \pm 0.26$    \\
\noalign{\smallskip} \hline\noalign{\smallskip}
 ~~ two gluon  exchange model  (this work)~~&  $3.85\pm 0.96$   \\
\noalign{\smallskip} \hline\noalign{\smallskip}
 effective  pomeron model (this work)    &  $3.75 \pm   0.84$ \\
\noalign{\smallskip}
\hline
\end{tabular}
\label{tab:jpsi-comparson}
\end{table}

  \begin{figure}[htbp]
\begin{center}
\includegraphics[scale=0.4]{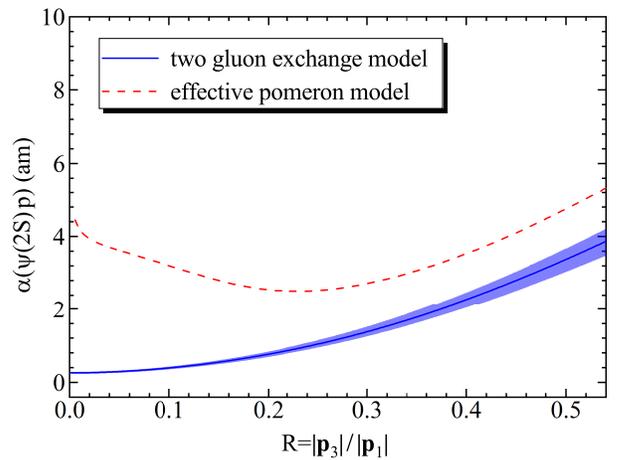}
\caption{ The obtained  scattering length  $\left|\alpha_{\psi(2S) p}\right|$ as a function of $R$.  The curves have the same meaning as in Fig.\ref{fig:psi}.
 } \label{fig:2ssl}
\end{center}
\end{figure}

\begin{table}
\caption{The   average value of scattering length     $\left|\alpha_{ \psi(2S) p}\right|$ obtained from  two gluon exchange model and effective pomeron model. }
\begin{tabular}{cc c}
\hline\hline\noalign{\smallskip}
  method  &  two gluon exchange model &  effective pomeron model  \\
\noalign{\smallskip} \hline\noalign{\smallskip}
  $ \left|\alpha_{ \psi(2S)  p}\right|$ (am) &  $ 1.31 \pm 0.92 $ & $ 3.24 \pm 0.63 $ \\
\noalign{\smallskip} \hline\noalign{\smallskip}
\end{tabular}
\label{tab:2s-sl}
\end{table}

\begin{table}[h]
\centering
\caption{  The values of  scattering length $\left|\alpha_{V p}\right|$   from  $\omega, \phi, ~ \text{and}~ \Upsilon $ meson.  }
\begin{tabular}{ccc }
\hline\hline\noalign{\smallskip}
~~~vector meson ~~~  &~~~ $m_V$ (GeV) ~~~&~~~~ $\left|\alpha_{V p}\right|$ (fm)~~~~\\
\noalign{\smallskip}\hline\noalign{\smallskip}
   $\omega $  & 0.87265 &       $0.82\pm 0.03$ \cite{Strakovsky:2014wja}    \\
\noalign{\smallskip}\hline\noalign{\smallskip}
   $\phi $  & 1.01946 &       $0.063\pm0.010$ \cite{Strakovsky:2020uqs}   \\
\noalign{\smallskip}\hline\noalign{\smallskip}
 $ \Upsilon $ & 9.4603 &  ~~  $(0.51\pm0.03)$ $\times 10^{-3}$  \cite{Strakovsky:2021vyk}~~\\
\noalign{\smallskip}
\hline
\end{tabular}
\label{tab:5}
\end{table}

\begin{figure}[htbp]
\begin{center}
\includegraphics[scale=0.4]{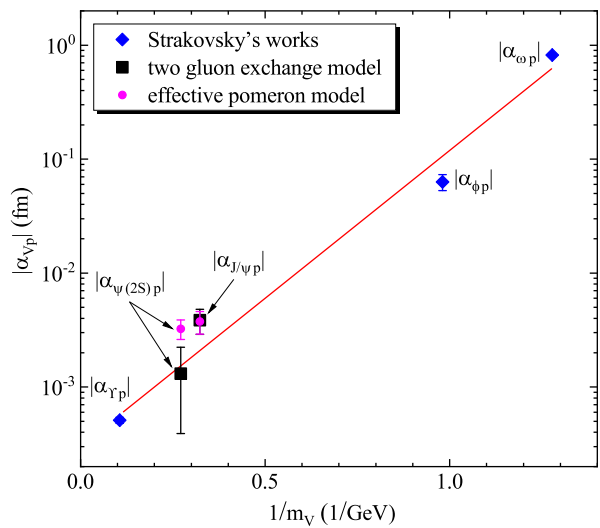}
\caption{Comparison of the scattering lengths  $\left|\alpha_{V p}\right|$  as a function of the inverse mass of  vector mesons,  including   $\omega, \phi, J/\psi,\psi(2S),$ and $ \Upsilon$. The blue-rhombus show the analysis of $\omega$ ( Strakovsky 2015:  \cite{Strakovsky:2014wja}), $\phi$ (Strakovsky 2020: \cite{Strakovsky:2020uqs}) and $\Upsilon$ ( Strakovsky 2021:  \cite{Strakovsky:2021vyk}).     The magenta-circles  show  the analysis  from two gluon exchange model, black-squares from effective pomeron model.
The red-solid line is hypothetical \cite{Strakovsky:2021vyk}. }
\label{fig:alpha}
\end{center}
\end{figure}

In the above study,  we found that the scattering lengths $\left|\alpha_{ J/\psi  p}\right|$   given by the two models are approximate and consistent with other literatures \cite{Pentchev:2020kao,Strakovsky:2019bev}. The reliability of the two models and VMD method was determined.
   Next, we will extend the study to $\psi(2S)p$ scattering length based on the above study.
The photoproduction of    $\psi(2S)$ is obtained by using the same parameters    listed in Tab. \ref{tab:GluonParameters} and  \ref{tab:JPAC}.
 Note that, the  parameter $A_{\psi(2S)}$ in Eq. (\ref{eq:F})  can be written as $A_{\psi(2S)}=R_{\psi(2S)} \cdot A_{J/\psi }$, and the relative strength
$R_{\psi(2S)}=0.55$ is obtained from the extraction by CLEO \cite{CLEO:2009zjv,Albaladejo:2020tzt}.
 Then the  $ \psi(2S)   p$ scattering length as a function of $R$ is  shown in Fig.\ref{fig:2ssl} (Red-dashed curve).
We obtained  calculation results from  two gluon exchange model  and effective pomeron  model to be      $ 1.31  \pm 0.92  \text{ am} $   and $ 3.24 \pm 0.63     \text{ am}$, respectively (table \ref{tab:2s-sl}).
 Note that, the value of $\left|\alpha_{ J/\psi  p}\right|$  is bigger than $\left|\alpha_{ \psi(2S)  p}\right|$    in both models.
Because of the difference in the size of the total cross section predicted by the two models, the  $\left|\alpha_{ \psi(2S)  p}\right|$  scattering length is a little bit different. Precise results  require more measurement from subsequent experiments.

Based on the recent threshold measurements of the photoproduction of $ \omega$ and $ \phi $    mesons  off the proton by the A2 (MAMI) \cite{Strakovsky:2014wja} and CLAS (JLab) \cite{Dey:2014tfa}, one can determine vector meson proton scattering lengths $\left|\alpha_{V p}\right|$  using the VMD model \cite{Strakovsky:2014wja,Strakovsky:2020uqs}. What is more,  the  absolute value of the  $\Upsilon   p$   scattering length is studied using quasi data generated from the QCD model \cite{Guo:2021ibg,Strakovsky:2021vyk}. The corresponding results for the scattering lengths are shown in table \ref{tab:5} and Fig.\ref{fig:alpha} as a function of the inverse vector  meson mass. Concretely,  the relationship including $ \omega, \phi,  J/\psi, \psi(2S), ~\text{and}~ \Upsilon$  can be determined as
\begin{align}
\left|\alpha_{\Upsilon p}\right| <  \left|\alpha_{\psi(2S) p}\right| <  \left|\alpha_{J/\psi p}\right| <  \left|\alpha_{\phi p}\right| <  \left|\alpha_{\omega p}\right| .
\end{align}
Actually, the binding energy $ E_b$ in nuclear matter can be determined by scattering length  $\left|\alpha_{V p}\right|$.
In a  linear density approximation, the binding energy $ E_b$ can be written as \cite{Kaidalov:1992hd}
\begin{align}
 E_b \backsimeq  \frac{2 \pi (M_N+M_{\psi(2S)}) \alpha_{\psi(2S)p}    }{  M_N M_{\psi(2S)}} \rho_{nm},
\end{align}
in which  the nuclear matter density $\rho_{nm}\backsimeq$  0.17 fm$^{-3}$.  Usually, we define the initial stage of meson formation as the ``young" stage of meson, and there is also a binding energy between meson and nucleon.   The   $\psi(2S) $ binding energy in  nuclear matter is calculated as $0.073 \text{ MeV}$.
Therefore, the smallness of the binding energy may be related to the ``Young" age of the vector mesons participating in the interaction with the proton. As a primitive meson, its properties show some differences.   The weak combination between $\psi(2S)$ and proton can promote the reaction of $ \psi(2S) p \rightarrow   \psi(2S) p  $.
Moreover, if it is assumed that the interaction between vector meson and nucleon needs more time to reach equilibrium specifically for slow heavy quarkoniums ($J/\psi$, $\psi(2S) $, and $\Upsilon$), this means that the ``Young" vector meson effect is more pronounced for heavy quarkoniums. For light vector mesons (such as $\omega$, $\phi$, etc.), this ``Young" vector meson effect may be relatively weak.

 \section{summary}\label{sec:summary}

In this paper,   the  value of the scattering length is expressed as a function of the ratio between  $\sigma (W)$ and $R (W)$ within the VMD model. This description can avoid  delivering unnecessary inaccuracy in numerical calculation. It is not only suitable for extracting the scattering length from the experimental data directly, but also convenient for observing the  near-threshold overall situation  for  theoretical model.
In this paper,  we research  the $J/\psi$ and $\psi(2S)$ scattering lengths  according to  the theoretical study of  charmoniums  photoproduction within  two gluon exchange model and effective pomeron model.
The  scattering lengths $\left|\alpha_{J/\psi p}\right|$ from the two models  are basically  consistent  with  several  other  theoretical predictions.
  Moreover,   the scattering length  of  $\psi(2S) $-proton interaction extracted from two gluon exchange model is $ 1.31\pm 0.92$ am. Additionally,  $\left|\alpha_{\psi(2S)p}\right|=3.24 \pm 0.63$ am   extracted from effective pomeron model.
The value of $\left|\alpha_{ J/\psi  p}\right|$  is bigger than $\left|\alpha_{ \psi(2S)  p}\right|$ in both models. And these results satisfy the nonlinear exponential increase  $\left|\alpha_{V p}\right| \propto  \text{exp}(1/m_V )$ basically.
According to the present results, it can be roughly concluded that one of the main factors affecting the scattering length  $\left|\alpha_{V p}\right|$ is the size of the corresponding cross section of vector meson photoproduction. For example, the cross section of the $\phi$ meson photoproduction is more than two orders of magnitude higher than the cross section of $J/\psi$ photoproduction, and correspondingly, the scattering length $\left|\alpha_{ \phi  p}\right|$ is nearly twenty times higher than the $\left|\alpha_{ J/\psi  p}\right|$. In addition, the cross section at the threshold is generally more complicated, and the results given by the two models also show differences. However, due to the lack of experimental data, especially the experimental data of $\psi(2S) $ photoproduction, the scattering length of $\psi(2S) $-proton cannot be determined very accurately.
Therefore, to better determine the scattering lengths of vector mesons and proton interaction, more high-precision experimental measurements for the photo/electro-production of charmoniums are highly needed, which can not only be realized in the JLab experiment \cite{GlueX:2019mkq}, but also within the capabilities of EicC and US-EIC facility \cite{Anderle:2021wcy,Accardi:2012qut}.

 \section{Acknowledgments}

X.-Y. Wang would like to acknowledge Dr. Daniel Winney for useful discussion about the  effective pomeron model.
This project is supported by the National Natural Science Foundation of China (Grant Nos. 12065014 and 12047501),
and by the West Light Foundation of The Chinese Academy of Sciences, Grant No. 21JR7RA201. IIS was supported in part by the US Department of Energy, Office of Science, Office of Nuclear Physics,   under Award No. DE-SC0016583.

\end{document}